\documentclass[review]{elsarticle}

\usepackage{hyperref}
\usepackage{graphicx}
\usepackage{dcolumn}
\usepackage{bm}
\usepackage[mathlines]{lineno}
\usepackage{xcolor}
\usepackage{xspace}
\usepackage{soul} 

\newcommand*{\pT}{\ensuremath{p_\mathrm{T}}\xspace}
\newcommand*{\pTl}{\ensuremath{p_\mathrm{T}^{\rm trigger}}\xspace}
\newcommand*{\pTa}{\ensuremath{p_\mathrm{T}^{\rm assoc.}}\xspace}

\newcommand*{\sumpTa}{\ensuremath{\sum p_\mathrm{T}^{\rm assoc.}}\xspace}

\newcommand*{\ncha}{\ensuremath{n_{\rm ch}^{\rm assoc.}}\xspace}
\newcommand*{\Nmpi}{\ensuremath{N_{\rm MPI}}\xspace}
\newcommand*{\avgNmpi}{\ensuremath{\left<N_{\rm MPI}\right>}\xspace}
\newcommand{\py}{\textsc{PYTHIA}\xspace}

\journal{Journal of \LaTeX\ Templates}

\begin{document}

\begin{frontmatter}

\title{Probing the interaction of semi-hard quarks and gluons with the underlying event in light- and heavy-flavor triggered proton-proton collisions}
\author[add1]{R\'obert V\'ertesi}
\author[add1]{Gyula Benc\'edi}
\ead{bencedi.gyula@wigner.hu}
\cortext[mycorrespondingauthor]{Corresponding author}
\author[add1,add2]{Anett Mis\'ak}%
\author[add3]{Antonio Ortiz}

\address[add1]{Wigner Research Centre for Physics, Hungary}
\address[add2]{Eötvös Loránd University Faculty of Sciences}
\address[add3]{Instituto de Ciencias Nucleares, Universidad Nacional Aut\'onoma de M\'exico}%

\date{\today}

\begin{abstract}
We study underlying-event observables in inelastic proton-proton (pp) collisions at a centre-of-mass energy of $\sqrt{s} = 13$\,TeV with identified light and heavy-flavor triggers using the \py~8 event generator. The study is performed as a function of the transverse momentum of the leading particle (\pTl). While at high \pTl ($>10$\,GeV/$c$) the underlying-event activity is independent of the leading particle species, at intermediate \pTl ($2<\pTl<8$\,GeV/$c$) it is larger in pion-triggered events than in events triggered with B mesons. Moreover, the underlying event in pion-triggered events, the majority of which are initiated by gluons, shows a stronger effect of color reconnection than events triggered with B-hadrons, that are mostly initiated by quark jets. The effect is observed at both hadronic and partonic level.  Given that color reconnection affects the interaction among final partons before the hadronization, and that in the string model quarks (gluons) are connected to one (two) string piece(s), we conclude that the observed effect can be attributed to differences in the interactions of gluon and quark jets with the underlying event.
\end{abstract}


\end{frontmatter}

\section{\label{sec:intro}Introduction}

Long-range multi-particle correlations in the final state have been considered as a signature of hydrodynamic evolution of the strongly coupled quark-gluon plasma (sQGP) \cite{Adcox:2004mh, Adams:2005dq, Back:2004je, Arsene:2004fa}.
LHC results, however, unexpectedly revealed that high-multiplicity proton-proton events exhibit similar collective behavior, with substantial azimuthal anisotropy ($v_n$) \cite{Khachatryan:2010gv,Acharya:2019vdf}.
Part of these effects can be explained with vacuum-QCD processes at the soft-hard boundary, such as multiple-parton interactions (MPI)~\cite{Bierlich:2018lbp}, although the creation of sQGP in a small volume cannot be completely ruled out~\cite{Bierlich:2017vhg,Blok:2017pui,Yan:2013laa,Trainor:2019mcp}. It is worth mentioning that the experimental data support the presence of MPI in high-energy hadronic interactions, see e.g. Refs.~\cite{Ortiz:2021peu,Ortiz:2020rwg}. In models that handle color flow in softer MPI processes in a simplified manner, an additional step is applied~\cite{Sjostrand:2017cdm}. This step, called the color reconnection (CR), has a major role in forming the multiplicity as well as the transverse momentum (\pT) distribution of the final state particles, and may be a key ingredient in the formation of collective behavior~\cite{Ortiz:2016kpz,Varga:2018isd}. Hence vacuum QCD effects establish connection between the leading hard process and the underlying event (UE) from particle production in soft and secondary hard processes as well as from beam remnants ~\cite{Ortiz:2013yxa}. The possibility of this connection resulting in the observed collectivity was explored in phenomenology calculations~\cite{Martin:2016igp,Ortiz:2018vgc,Ortiz:2017jaz} and later corroborated in the experiment~\cite{Acharya:2019nqn,Tripathy:2020lla}. These works, however, mostly concentrated on light-flavor and strangeness observables.
Jet fragmentation is affected by the flavor of the initial parton in several ways. First, light-flavor jets are initiated either by a gluon or a quark, while heavy-flavor jets are mostly initiated by quarks, therefore the average color charge of the initiating parton is different. Then, the heavy-quark fragmentation functions differ from those of light quarks and gluons~\cite{Dai:2018ywt}: heavy quark fragmentation is harder on the average than light-quark fragmentation, meaning that a larger part of the momentum is taken away by the heavy quarks~\cite{Vertesi:2020gqr}. Finally, heavy-flavor jet shapes at low \pT are influenced by the dead cone effect, or the suppression of forward gluon emissions~\cite{Dokshitzer:1991fd}. 

In this work, we use \py~8 simulations to explore the connection between the underlying event and the jet-initiating parton, using identified heavy and light-flavor triggers. We argue that the connection of the underlying event to the leading process depends on whether the initiating parton is a quark or a gluon, and flavor-dependent studies are an excellent experimental tool to understand this dependency. The article is organised as follows:  Section~\ref{sec:ana} provides information about the analysis approach, as well as the simulations using the \py~8 Monte Carlo generator. Section~\ref{sec:result} presents the results and discussion, and finally Section~\ref{sec:concl} summarizes the results.

\section{\label{sec:ana}Analysis method}

In our simulations we used \py~8.240~\cite{Sjostrand:2007gs} with the Monash 2013 tune~\cite{Skands:2014pea} and soft QCD settings. \py is an event generator that models a basic hard scattering process with leading-order pQCD calculations, amended with initial-state radiation (ISR), final-state radiation (FSR) to account for higher-order processes. Multiple-parton interactions are also handled on the partonic level, and CR is also applied at a later step. The hadronic final state is produced using Lund string fragmentation, and then secondary decays and rescattering between hadrons are computed.  The Monash 2013 tune is mainly focused on describing the minimum-bias and UE distributions accurately.
We simulated 100 million proton-proton events at $\sqrt{s}$ = 13 TeV with the following settings.

To characterize the events by underlying-event activity, in each event we identified the highest transverse momentum particle in the central pseudorapidity window $|\eta| < 0.8$ corresponding to the acceptance of the main tracking detector of the ALICE experiment~\cite{Cortese:2005qfz}. Only triggers above $\pTl=0.5$ GeV/$c$ were considered. We categorized events based on the trigger particles: as for heavy-flavor, we collected events where the triggers were either D-mesons ($\mathrm{D}^0$, $\mathrm{\bar{D}}^0$ and $\mathrm{D}^\pm$) or B-mesons ($\mathrm{B}^0$, $\mathrm{\bar{B}}^0$ and $\mathrm{B}^\pm$). For comparison, we identified charged pions ($\pi^\pm$) to represent light flavor. Instead of a reconstruction of heavy-flavor hadrons from the final state, we disabled their decays. This way we also avoid accounting for the feed-down contribution from B to D. Excited states, on the other hand, were allowed to decay, to simplify the analogous experimental environment.
Within each event we correlated all associated particles to the trigger that had a minimal transverse momentum of \pTa = 0.5 GeV/$c$, at central pseudorapidity (\begin{math} |\eta| < 0.8 \end{math}). We applied a spatial division based on the azimuth angle ($\Delta\varphi$) between the trigger and the associated particles. 
Following the definition by the CDF Collaboration, the direction of the trigger particle is used to define regions in the azimuthal plane that have different sensitivity to the UE~\cite{Buttar:2005gdq}. The azimuthal angular difference between the trigger particles and the associated particles is used to define the following azimuthal regions considered in this study: $|\Delta\varphi| < \frac{\pi}{3}$ for the toward region and $\frac{\pi}{3} < |\Delta\varphi| < \frac{2\pi}{3}$ for the transverse region, dominated by the UE.
\py~8 is known to reproduce the charged-hadron multiplicities in the jet as well as in the underlying-event regions~\cite{Acharya:2019nqn}. Besides our reference simulations we also generated samples with non-physical settings, to understand the individual effect of model components: without FSR and ISR, disabled and enhanced CR.

\section{\label{sec:result}Results and Discussion}

\begin{figure}[t]
\centering
\includegraphics[width=1\columnwidth, keepaspectratio]{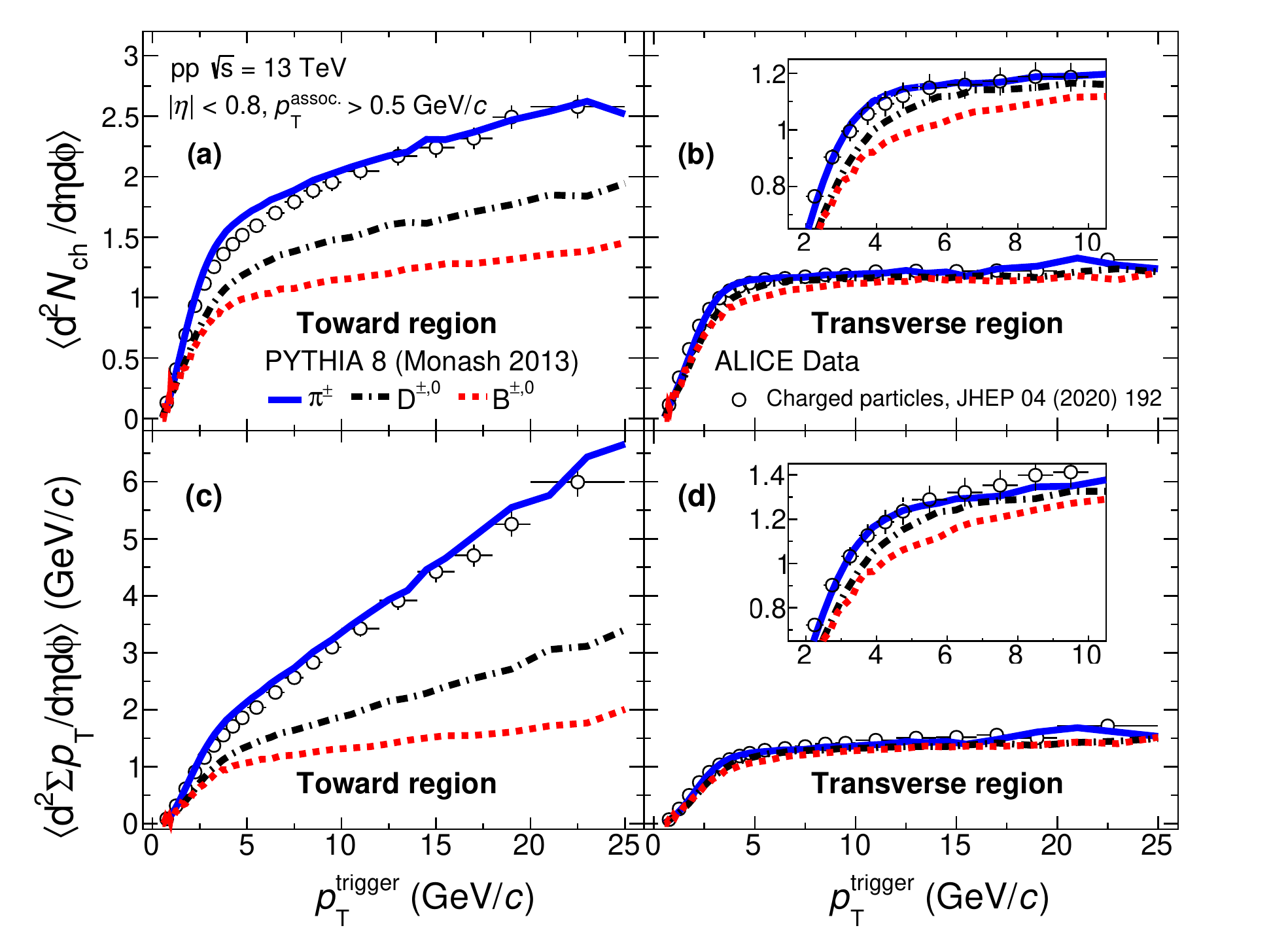}%
\begin{center}
    \caption{\label{fig:NchPt} 
    Average charged-particle number (panels (a) and (b)) and summed-\pT (panels (c) and (d)) densities as a function of \pTl for charged pions, D- and B-meson triggers in the toward (left column) and transverse (right column) regions.
    The \py~8 predictions are compared to experimental data (open circles) for charged particles measured by the ALICE Collaboration~\cite{Acharya:2019nqn}.
    }
\end{center}
\end{figure}

We characterized the particle production both with the average density of associated particles  $\langle\mathrm{d}^{2}N_{\rm ch}/\mathrm{d}\eta\mathrm{d}\varphi\rangle\equiv n_{\rm ch}^{\rm assoc.}$, where $N_{\rm ch}$ is the number of particles in the selected region, and the average transverse momentum sum density $\langle\mathrm{d}^{2}\sum{p}_{\rm T}/\mathrm{d}\eta\mathrm{d}\varphi\rangle\equiv\sumpTa$ carried by these particles. These distributions are shown in Fig.~\ref{fig:NchPt} for both the toward and the transverse regions, and they are compared to recent experimental data (open circles) for charged particles measured by the ALICE Collaboration~\cite{Acharya:2019nqn}.
In the toward region, a monotonic rise of both \ncha and \sumpTa is present with \pTl. In case of \sumpTa this rise is approximately linear, revealing that the trigger particle carries an approximately constant fraction of the full jet momentum in the observed momentum range. Since higher-momentum jets fragment into particles with higher average \pT, the steepness of the \ncha curve decreases toward higher \pTl.
The ALICE data are well described by simulations in which charged pions are the trigger particles. However, if one considers heavy-flavour triggered events, the activity is significantly reduced. This remarkable flavor hierarchy represents the flavor (mass) dependent fragmentation.
Similarly to what has been observed for light flavor~\cite{Acharya:2019nqn}, in the transverse region there is a near-saturation towards high \pTl. We see that the saturation value does not depend significantly on the flavor of the trigger. This indicates that the leading process is largely independent of the UE. However, in the intermediate transverse-momentum range $2<\pTl<8$ GeV/$c$ the transverse event activity is different for each flavor: the underlying event activity in pion-triggered events is higher than that for heavy-flavor triggers. The effect is similar, albeit much larger, than what has been found for different light-flavor hadrons~\cite{Ortiz:2018vgc}, which shows that the effect cannot be explained by differences in the the mass of the (final-state) trigger hadron. In the same study, however, color reconnection (CR) was found to be the reason for the hadron-quality-dependent behavior.

\begin{figure}[t]
\centering
\includegraphics[width=1\columnwidth, keepaspectratio]{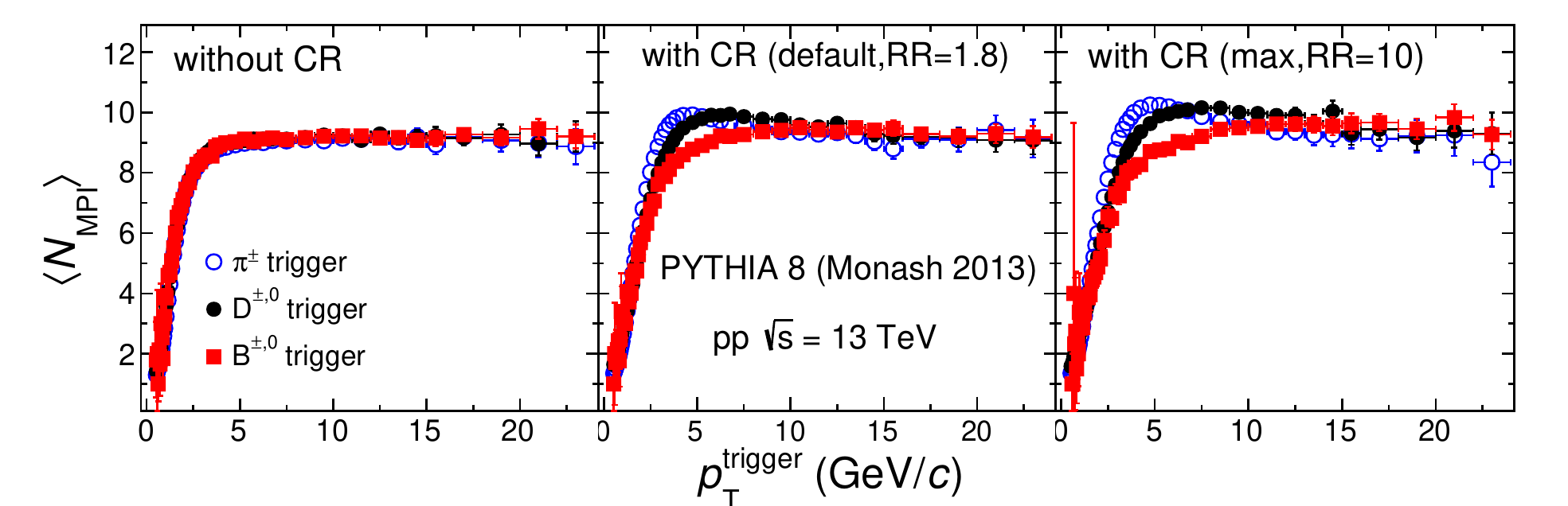}%
\begin{center}
    \caption{\label{fig:Nmpi} 
    Average number of multiple-parton interactions as a function of \pTl for charged pions, D- and B-meson triggers without and with color reconnection. Panels show different values of color reconnection range ($RR$): $RR=0$, $RR=1.8$, $RR=10$ in the left, middle and right panels, respectively.
     }
\end{center}
\end{figure}

The event activity in the transverse region is related to the number of multi-parton interactions, \Nmpi, in an event~\cite{Martin:2016igp}. In order to gain a deeper insight, we investigate \avgNmpi in simulations with different strengths of color reconnection, controlled in \py with the Reconnection Range parameter ($RR$). Figure~\ref{fig:Nmpi} shows the average number of multiple-parton interactions per trigger, \avgNmpi, as a  function of \pTl, for the different identified triggers. 
We see that if there is no color reconnection ($RR=0$), then there is no significant difference between \Nmpi for any of the trigger hadron flavors. If the color reconnection matches the physical tune ($RR=1.8$), a clear difference can be observed in \avgNmpi depending on the trigger particle species in the momentum range $2<\pTl<8$ GeV/$c$, similarly to what is present for \ncha and \sumpTa. The MPI activity corresponding to light-flavor triggers is enhanced in this region compared to heavy-flavor triggered events, forming a visible bump structure. This bump is already much less prominent in the case of D-mesons, and for B-mesons the MPI activity grows monotonically until saturation. As can be expected, the effect is even more pronounced in the enhanced color-reconnection scenario. It is noteworthy that in the toward region the flavor ordering is present even if there is no color reconnection. Therefore, it can be assumed that the ordering in the transverse region arises in connection with the ordering observed in the jet region, via final-state collective traits produced by CR~\cite{Ortiz:2016kpz}.

In the string fragmentation model of \py, the main difference between quark and gluon hadronization is that gluons are connected to two string pieces, while quarks are only connected to a single string piece~\cite{Sjostrand:2006za}. One would expect that selecting gluon jets biases the sample towards events with a larger number of strings than selecting enhanced quark jet samples. Given the larger number of strings in events with gluon jets than with quark jets, color reconnection is expected to give a larger flow-like effect if enhanced gluon jet samples are analyzed. To understand whether what we observed using pion or B meson trigger particles is connected with this, we estimate the fraction of gluon (quark) jets in triggered leading-pion (-B meson) events.

\begin{figure}[t]
\centering
\includegraphics[width=1\columnwidth,keepaspectratio]{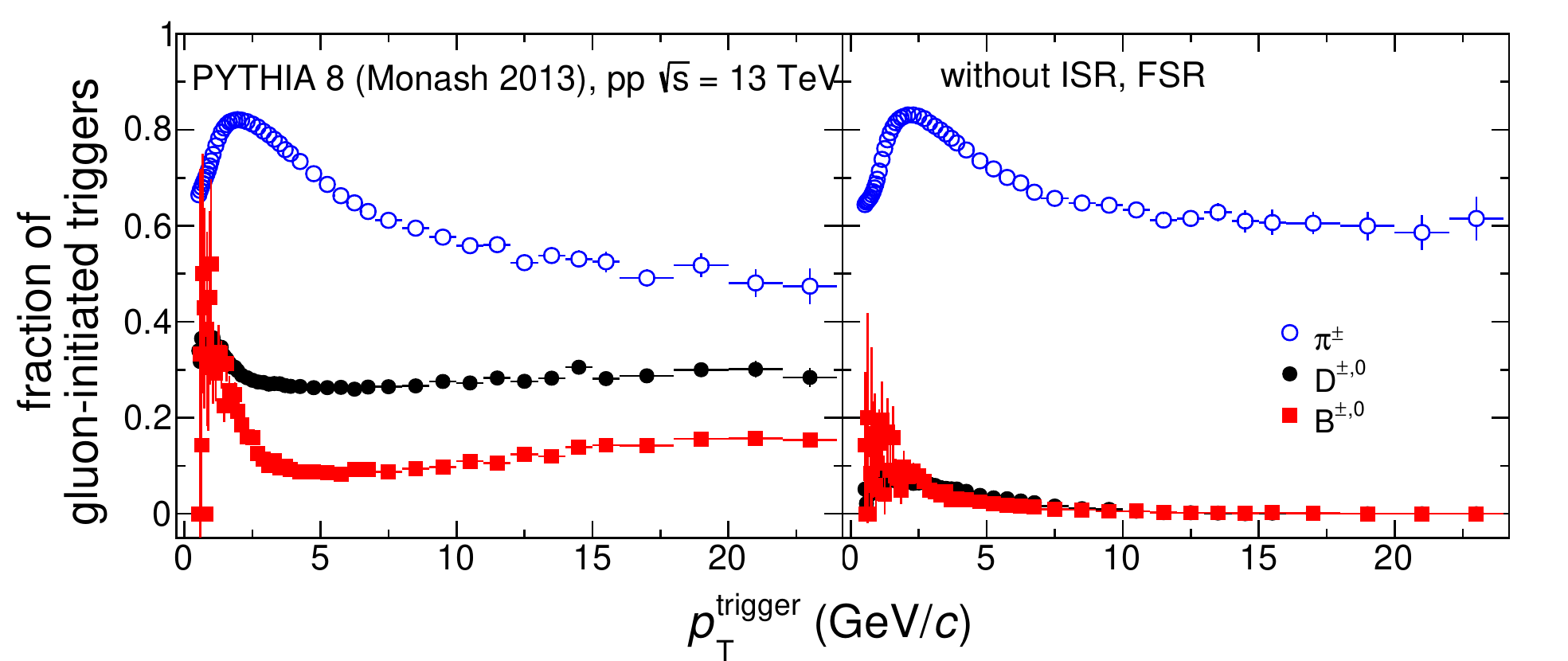}%
\begin{center}
    \caption{\label{fig:ratioGl} Ratio of \pTl distributions with and without the selection on gluon-initiated triggers for charged pions, D-, and B-mesons. The left and the right panel shows the cases with and without ISR and FSR.
    }
\end{center}
\end{figure}

In order to estimate the gluon (quark) jet contribution in event samples containing a pion (B meson) trigger particle, we implemented an algorithm that identifies events based on whether the parton that eventually fragments into the trigger particle is a quark or a gluon. The initiating parton is not necessarily well-defined in a complex interaction. In \py there is no a priori way to determine the initiating parton type with full certainty. As an approximation, we applied the following method: we look for a high-\pT parton in a cone region around the trigger particle, with a radius $R=\sqrt{\Delta\varphi^2+\Delta\eta^2} < 0.5$. Since the trigger particle is most likely a jet constituent and it is expected to be close to the jet axis, this region is sensitive to jet fragmentation. The value of $R$ ensures that the dominant contribution in this region comes from the jet, and the jet will be mostly contained in this cone~\cite{Chatrchyan:2012mec}. To prevent the association of soft partons to the trigger particle, the presence of a high-\pT parton with a \pT of at least $30\%$ of the \pT of the trigger particle was required. Moreover, the selected parton should be marked ``ancestor'' by \py. If no such parton is found, then we look for the closest parton in the event record that has $\pT>0.3\cdot\pTl$ and is within the $R<0.5$ cone. If no such parton exists, then we finally look for the closest parton in the event record that is marked ``ancestor'' by \py. As a next step we assess the effectivity of our sorting algorithm. Figure~\ref{fig:ratioGl} shows the fractions of charged pion, D-meson and B-meson triggers that are identified as gluon-initiated to the case without the selection of the initiating parton.
The left panel shows the physical case, while in the right panel the case is shown where initial-state radiation (ISR) and final-state radiation (FSR) are turned off. Since in this case no radiated gluons are present, we expect that our selection algorithm works reliably. In the case where ISR and FSR are allowed, the algorithm identifies up to 30\% of D-meson triggers and 10--15\% of B-meson triggers as gluon-initiated. On the other hand, without radiation contribution the fraction of heavy-flavor triggers identified as gluon-initiated is below 5--10\% at low-\pTl and vanishes towards higher \pTl.
Regarding pion triggers, the algorithm identifies more than 60\% of the events as gluon-initiated in the $0.5<\pTl<5$ GeV/$c$ interval. As a conclusion, B-jets can be used as proxies for quark jets within an error margin below 5--10\%, and pions can be used as proxies for gluon jets albeit with a higher misidentification rate, between 10-25\% depending on the observed \pTl range. It is to be noted that higher-order hard processes, such as gluon splitting and flavor excitation, comprise a significant contribution to heavy flavor production~\cite{Norrbin:2000zc}. Therefore the parton that initiates the triggering jet cannot be directly identified to the hardest pQCD process. Nevertheless, we also cross-checked our algorithm with \py where gluon and quark processes were explicitly selected first as a \py hard QCD process, and also as the first scattering process in soft QCD. We found similar patterns in both cases to our initiating parton selection algorithm. However, a larger fraction (up to 35\%) of heavy-flavor triggers were recorded to come from events with a gluon leading process. 

\begin{figure}[t]
\centering
\includegraphics[width=1\columnwidth, keepaspectratio]{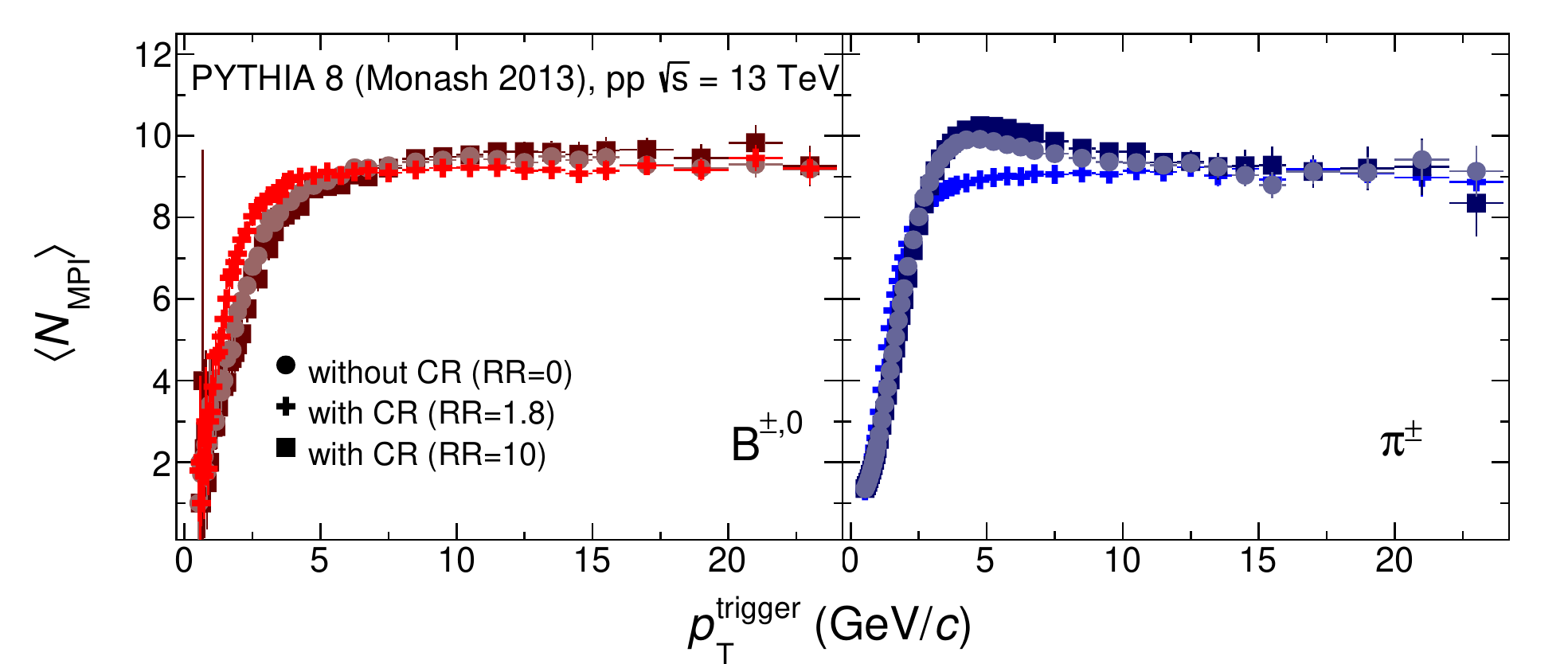}%
\begin{center}
    \caption{\label{fig:NmpiBpiCR} Average number of multiple-parton interactions for B-meson (left panel) and charged pion (right panel) triggers for different values of color reconnection ranges ($RR$): $RR=0$, $RR=1.8$, and $RR=10$.
    }
\end{center}
\end{figure}

In order to see how well the flavor separation might work in an experimental environment, we compare the behavior of B-mesons to pions, as proxies for quark-initiated and gluon-initiated jets, for different CR strengths. In Fig.~\ref{fig:NmpiBpiCR} we show \avgNmpi as a function of \pTl for B-mesons and charged pions. For pion-triggered events, the increase of the color reconnection range causes a depletion of the \avgNmpi at low \pTl ($\pTl\approx 3$\,GeV/$c$) and an enhancement at high \pTl ($\pTl\approx 5$\,GeV/$c$). The bump is not observed for B-meson triggered events.

\begin{figure}[t]
\centering
\includegraphics[width=1\columnwidth, keepaspectratio]{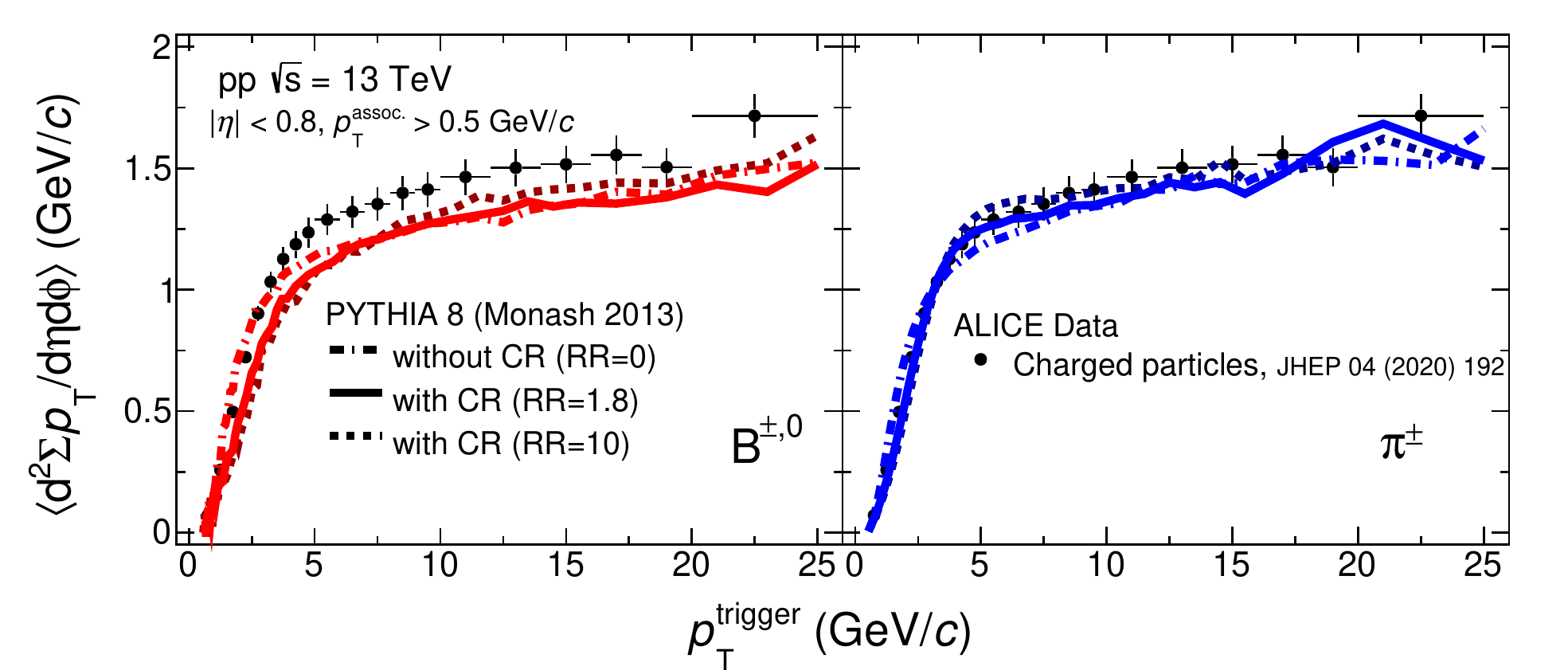}%
\begin{center}
    \caption{\label{fig:PtBpiCR} 
    Average summed-\pT densities for B-meson (left panel) and charged pion (right panel) triggers for different values of color reconnection ranges ($RR$): $RR=0$, $RR=1.8$, and $RR=10$. The \py~8 predictions are compared to experimental data (open circles) for charged particles measured by the ALICE Collaboration~\cite{Acharya:2019nqn}.
    }
\end{center}
\end{figure}

For the purpose of providing an experimentally accessible quantity, we show \sumpTa for different CR settings in a similar manner in Fig.~\ref{fig:PtBpiCR}. Our simulation results are also compared to recent experimental data (full circles) for charged particles measured by the ALICE Collaboration~\cite{Acharya:2019nqn}.
Note that while the pattern seen in \avgNmpi (in Fig.~\ref{fig:Nmpi}) is somewhat better represented in the transverse particle density than in the sum transverse momentum (shown in Fig.~\ref{fig:NchPt}), the presence of color reconnection is known to affect event multiplicity distributions~\cite{Varga:2018isd} and therefore the number of particles cannot be used when comparing different CR settings, while the momentum is conserved. 
While in the case of summed-\pT a weak but steady rise is present in the range where \avgNmpi saturates, the observed trends are again very similar. There is no significant difference above the point of onset for the $\pi^\pm$ case, while in the semi-hard \pTl range the summed-\pT distributions of light $\pi^\pm$ mesons show a characteristic dependence on CR. These observations corroborate the assumption that the leading hard process is connected to the UE via color reconnection and that the flavor quality of the leading process has a strong influence on the development of the UE. The observed pattern suggests that the flavor dependence is driven in a large part by the color composition of the initiating parton.
Experimental confirmation of the results presented here can be useful to improve our knowledge on the particle production mechanism in pp interactions, and their connection with the heavy-ion-like collective effects discovered at the LHC. 

\section{\label{sec:concl}Conclusions}

We investigated the particle production in the underlying event corresponding to identified light and heavy-flavor triggers, using \py~8 simulations. The particle yield in the transverse region, as well as the number of multiple-parton interactions that is strongly correlated to it, saturates towards higher \pTl, independently of particle species. However, as a consequence of color reconnection, flavor hierarchy shows up in the region $\pTl<8$ GeV/$c$ where soft processes play an important role. It can be assumed that this arises from the jet region being linked to the transverse region via collectivity generated by color reconnection. In the case of light flavor, \avgNmpi overshoots the saturation value, while this is not seen in heavy flavor. This ``bump'' structure observed at $\pTl\approx 4$ GeV/$c$ can be associated with gluon-initiated jets. We find that while pion triggers are mostly gluon-initiated, B-mesons serve as a good proxy for quark jets. This provides an opportunity to evaluate the connection of gluon- and quark-initiated jets to the underlying event by comparing the particle production in the transverse region by light-hadron and B-meson triggered events in the experiment, and thus gives new means for model development.

\section*{Acknowledgement}
This work was supported by the Hungarian National Research, Development and Innovation Office (NKFIH) under the contract numbers OTKA FK131979 and K135515, and the NKFIH grant 2019-2.1.11-T\'ET-2019-00078.  
A.~Ortiz acknowledges the support from CONACyT under the Grants CB No. A1-S-22917 and CF No. 2042.
The authors also acknowledge the computational resources provided by the Wigner GPU Laboratory and research infrastructure provided by the E\"otv\"os Lor\'and Research Network (ELKH).

\section*{References}

\bibliography{mybibfile}

\end{document}